\documentclass[acus]{JAC2003}

\usepackage{amsmath}
\usepackage{graphicx}
\usepackage{booktabs}
\usepackage{epstopdf}
\usepackage{balance}

\begin{document}
\title{BENCHMARKING HEADTAIL WITH ELECTRON CLOUD \\ INSTABILITIES OBSERVED IN THE LHC }

\author{H.~Bartosik, W.~H\"ofle, G.~Iadarola, Y.~Papaphilippou, G.~Rumolo \\
CERN, Geneva, Switzerland}

\maketitle

\begin{abstract}

After a successful scrubbing run in the beginning of 2011, the LHC can be presently operated with high intensity proton beams with 50\,ns bunch spacing. However, strong electron cloud effects were observed during machine studies with the nominal beam with 25\,ns bunch spacing. In particular, fast transverse instabilities were observed when attempting to inject trains of 48 bunches into the LHC for the first time. An analysis of the turn-by-turn bunch-by-bunch data from the transverse damper pick-ups during these injection studies is presented, showing a clear signature of the electron cloud effect. These experimental observations are reproduced using numerical simulations: the electron distribution before each bunch passage is generated with PyECLOUD and used as input for a set of HEADTAIL simulations. This paper describes the simulation method as well as the sensitivity of the results to the initial conditions for the electron build-up. The potential of this type of simulations and their clear limitations on the other hand are discussed.

\end{abstract}

 \section{INTRODUCTION}

\begin{figure}[h]
    \centering
    \includegraphics[trim= 5 0 30 0, clip,width=0.49\textwidth]{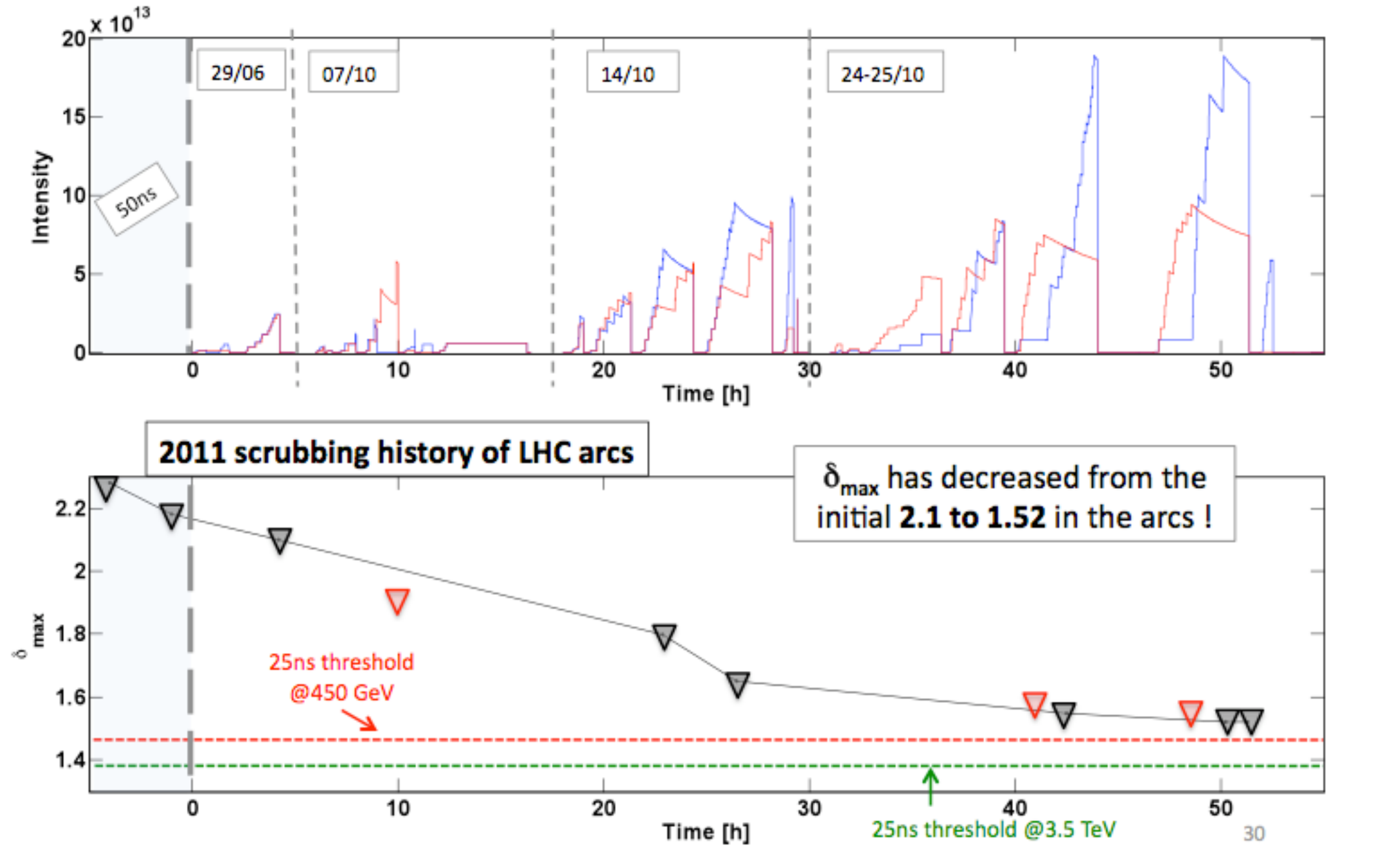}
     \caption{Evolution of $\delta_{max}$ in the LHC arcs as obtained by reproducing the measured heat load on the LHC beam screens using PyECLOUD \cite{GrumoloChamonix2012}.} 
    \label{FIG:EvolutionSEY}
\end{figure}

At the early phase of the 2011 LHC run, seven days were devoted to scrubbing using the 50\,ns bunch spacing beam with gradually increasing number of bunches circulating in the machine. This allowed to sufficiently condition the inner surface of the LHC beam screens and vacuum chambers for running with the same beam for physics production in routine operation throughout 2011. First attempts to inject a beam with the nominal 25\,ns bunch spacing were performed at the end of June 2011 with bunch trains of 24 bunches. Pressure rise and increased heat load in the arcs were observed. The studies continued with injections of 48 bunches in August, where the beams became transversely unstable after about 1000 turns with the transverse damper switched on and after about 500 turns without transverse damper, for two injections respectively. As will be discussed in more detail below, an analysis of the turn-by-turn bunch-by-bunch data from the transverse damper pick-ups~\cite{25nsMDnote} points to the observation of coherent electron cloud instabilities. During these injection tests the chromaticity was set to about $Q'\approx2$ in both planes, as usually used on the LHC flat bottom in routine operation. Only after increasing the chromaticity to about $Q'\approx15$ in both planes during further studies in October, it was possible to perform nominal injections of 288 bunches with 25\,ns bunch spacing from the SPS. The high chromaticity suppressed the fast instabilities observed before. However, the beam suffered from slow losses and transverse emittance blow-up along the bunch train and, as before, pressure rise and increased heat load were measured in the cold arcs. 
The conditioning of the LHC beam screens due to the beam based electron bombardment was demonstrated by estimating the secondary electron yield (SEY) from a comparison of the measured heat load data in the arcs with PyECLOUD simulations \cite{GiadarolECLOUD12,GrumoloEvian2011}. Figure~\ref{FIG:EvolutionSEY} shows the evolution of the obtained maximum SEY ($\delta_\text{max}$) as a function of beam time in the LHC together with the total intensity for both beams \cite{GrumoloChamonix2012}. Since the heat load data can be read only per half cell in the arcs, the SEY can be estimated only for both beams at the same time (grey markers) unless there is only one beam present in the machine (red markers for beam 2). A clear conditioning effect from $\delta_\text{max}\!=\!2.1$ at the end of June when running with 50\,ns bunch spacing to $\delta_\text{max}\!=\!1.52$ after approximately 50\,h beam time with 25\,ns bunch spacing is observed. The analysis and simulation studies presented in this paper will concentrate on the aforementioned injection tests with 48 bunches in August, where $\delta_{max}$ was estimated to be still around 2.1 (cf.~Fig.\,\ref{FIG:EvolutionSEY}).

\section{Analysis of turn-by-turn data}

\begin{figure*}[t]
    \centering
    \includegraphics[trim= 0 -1 0 0, clip,width=0.43\textwidth]{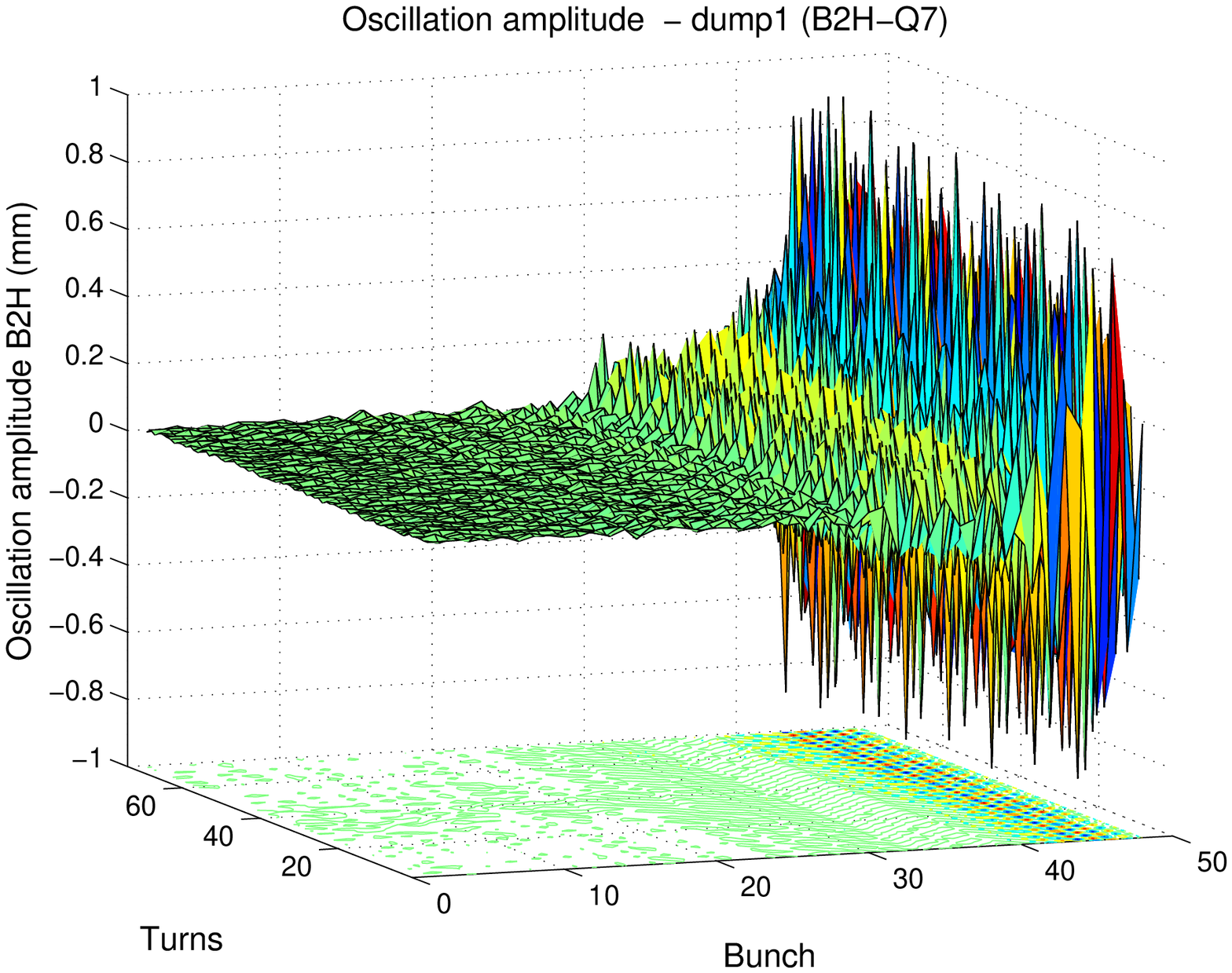}
    \includegraphics[trim= 0 -1 0 0, clip,width=0.43\textwidth]{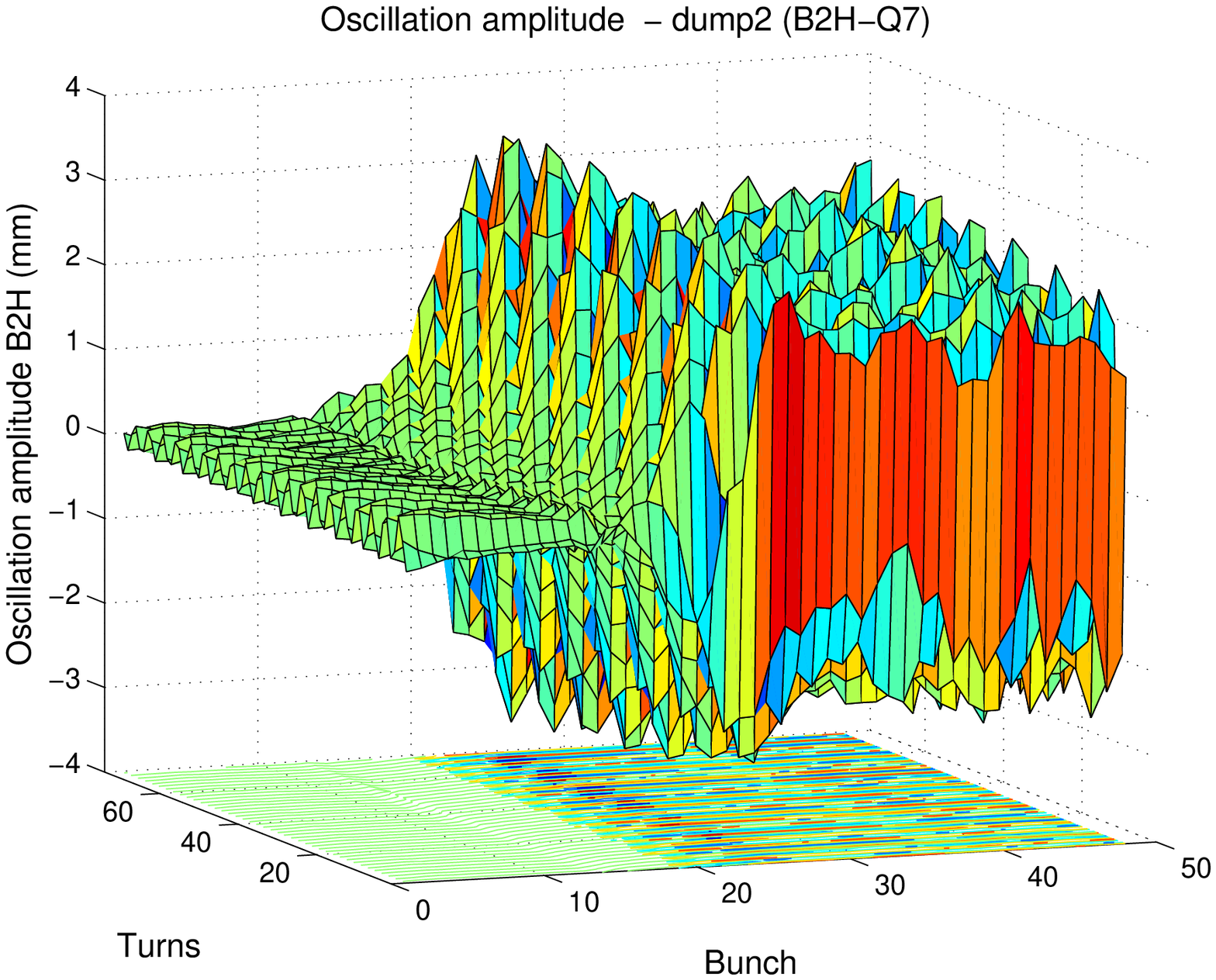}
    \includegraphics[trim= 0 0 0 -0, clip,width=0.43\textwidth]{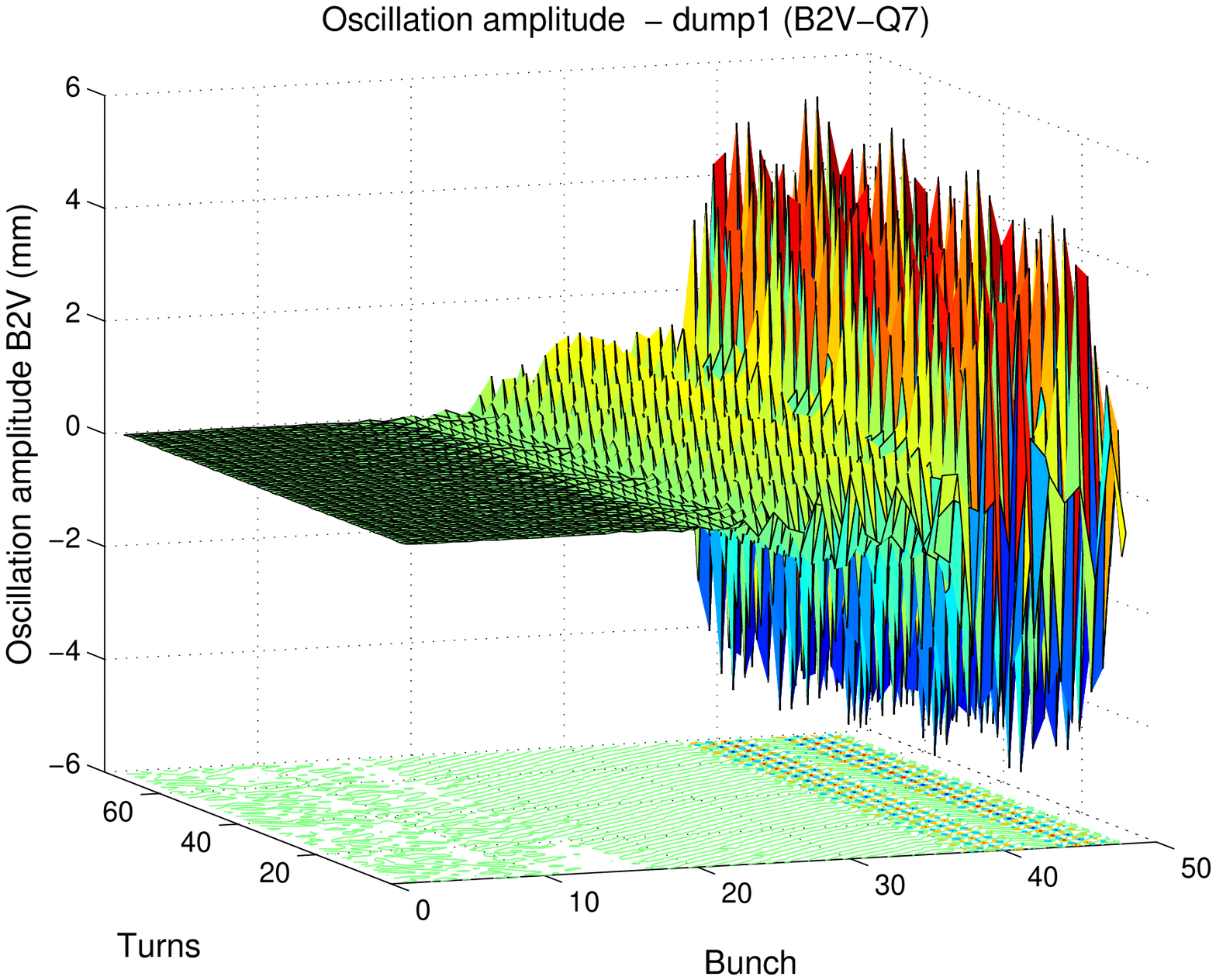}    
    \includegraphics[trim= 0 0 0 -0, clip,width=0.43\textwidth]{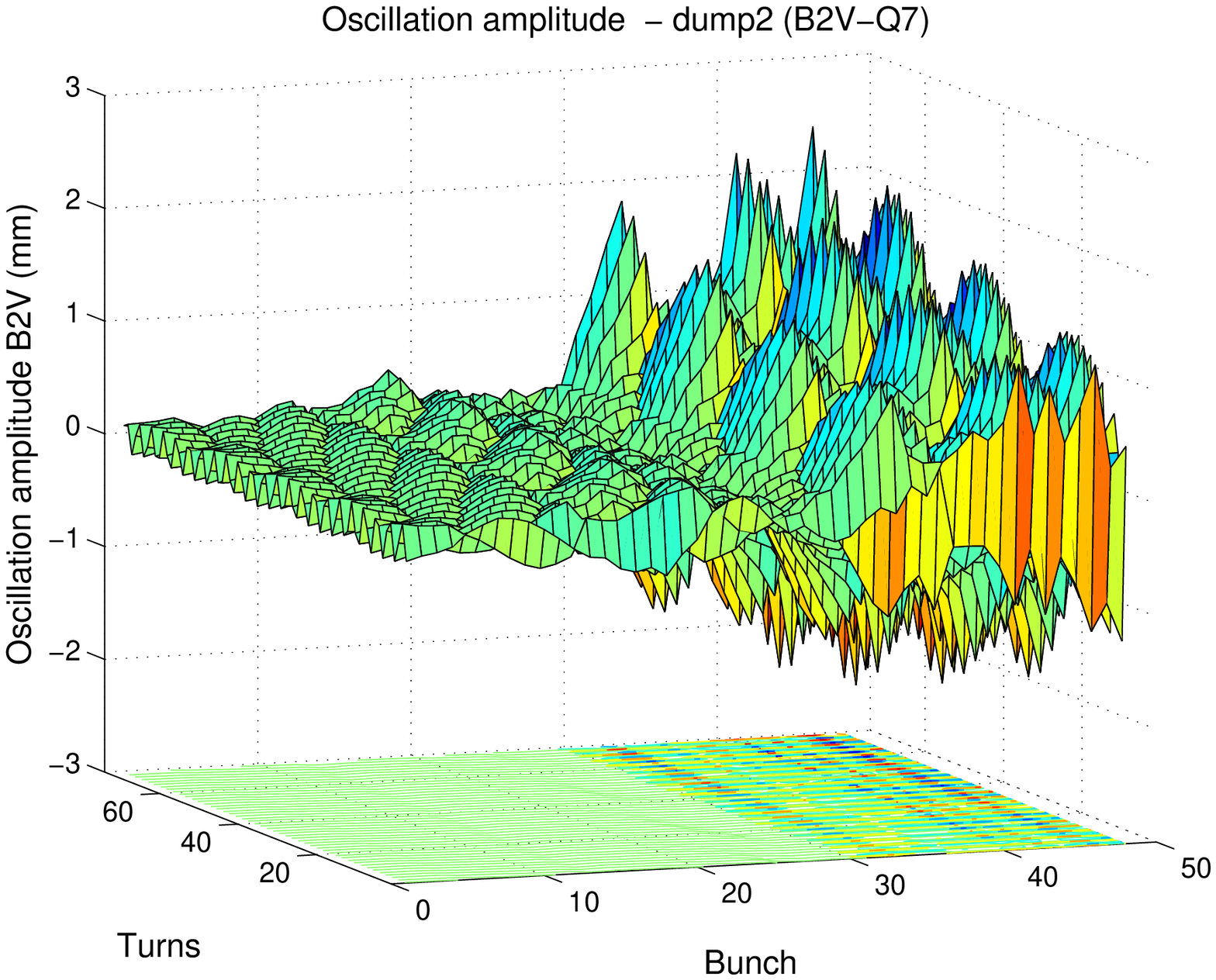}        
    \caption{Transverse oscillations measured with the pick-ups of the transverse feedback for 73 turns close before the beam dump, for the first case with transverse damper (left) and the second case without transverse damper (right) in the horizontal (top) and vertical plane (bottom).} 
    \label{FIG:ADToscillations}
\end{figure*}

As mentioned before, a first attempt to inject bunch trains of 48 bunches spaced by 25 ns into the LHC with the an intensity of about $1.0\!\times\!10^{11}$ p/b was made for beam 2 during a machine development session on August 26, 2011. Shorter bunch trains of 12 and 24 bunches with the same bunch spacing had been successfully injected earlier in the session. It was planned to establish the injection of longer bunch trains and complete the transverse damper set-up for the nominal bunch spacing of 25\,ns. A first injection of 48 bunches with the transverse damper switched on was dumped after 1000 turns due to a beam excursion interlock (referred to as dump 1). Another injection of 48 bunches without transverse damper was aborted already after around 500 turns due to a beam loss interlock (referred to as dump 2). In both cases, chromaticity was set to the values usually used during operation ($Q'\approx2$). 

Data of the damper pick-ups for about 73 turns are stored for both beams in the post mortem system at a beam abort. Each data set (channel) represents one of the allowed 3564 bunch positions per ring. Figure~\ref{FIG:ADToscillations} shows the turn-by-turn oscillation of each of the 48 bunches around the closed orbit in both planes for the 2 injections just before the respective beam dump. The oscillation amplitude is very small for the first 25 bunches, especially when the damper is switched on. In this case the last bunches of the bunch train reach peak values of around 1\,mm in the horizontal plane and up to 6\,mm in the vertical plane, thus the instability is mainly observed in the vertical plane. Without damper, the last bunches of the train exhibit oscillations of up to 3\,mm amplitude in both planes. These observations are compatible with a coupled bunch instability (slightly stronger in the horizontal plane than in the vertical plane), which can be suppressed by the damper, and a high frequency instability with a broad spectrum mainly in the vertical plane. The frequency spectrum without damper is dominated by coupled bunch modes up to 1\,MHz in the horizontal plane and up to 2\,MHz in the vertical plane. With damper on, instabilities in both planes are damped up to a frequency of about 15\,MHz~\cite{25nsMDnote}. This can be explained by a ``single bunch'' instability in the vertical plane and a coupled bunch instability (mainly) in the horizontal plane, as has been observed in the past already in the SPS~\cite{Gianluigi}. The beam observations during the injection tests in the LHC are used to benchmark numerical simulations as described in the following sections.

\section{Simulation procedure}

\begin{figure}[ht]
    \centering
    \includegraphics[trim= 0 0 0 0, clip,width=0.43\textwidth]{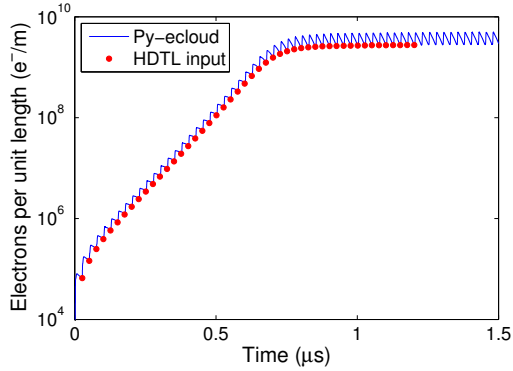}
    \caption{Electron density per unit length along the bunch train as obtained from a PyECLOUD simulation. A snapshot of the electron distribution before each bunch entrance as indicated by the red dots is used as input for individual HEADTAIL simulations.} 
    \label{FIG:PyECLOUDbuildup}
\end{figure}

\begin{figure}[h]
    \centering
    \includegraphics[trim= 0 0 0 0, clip,width=0.42\textwidth]{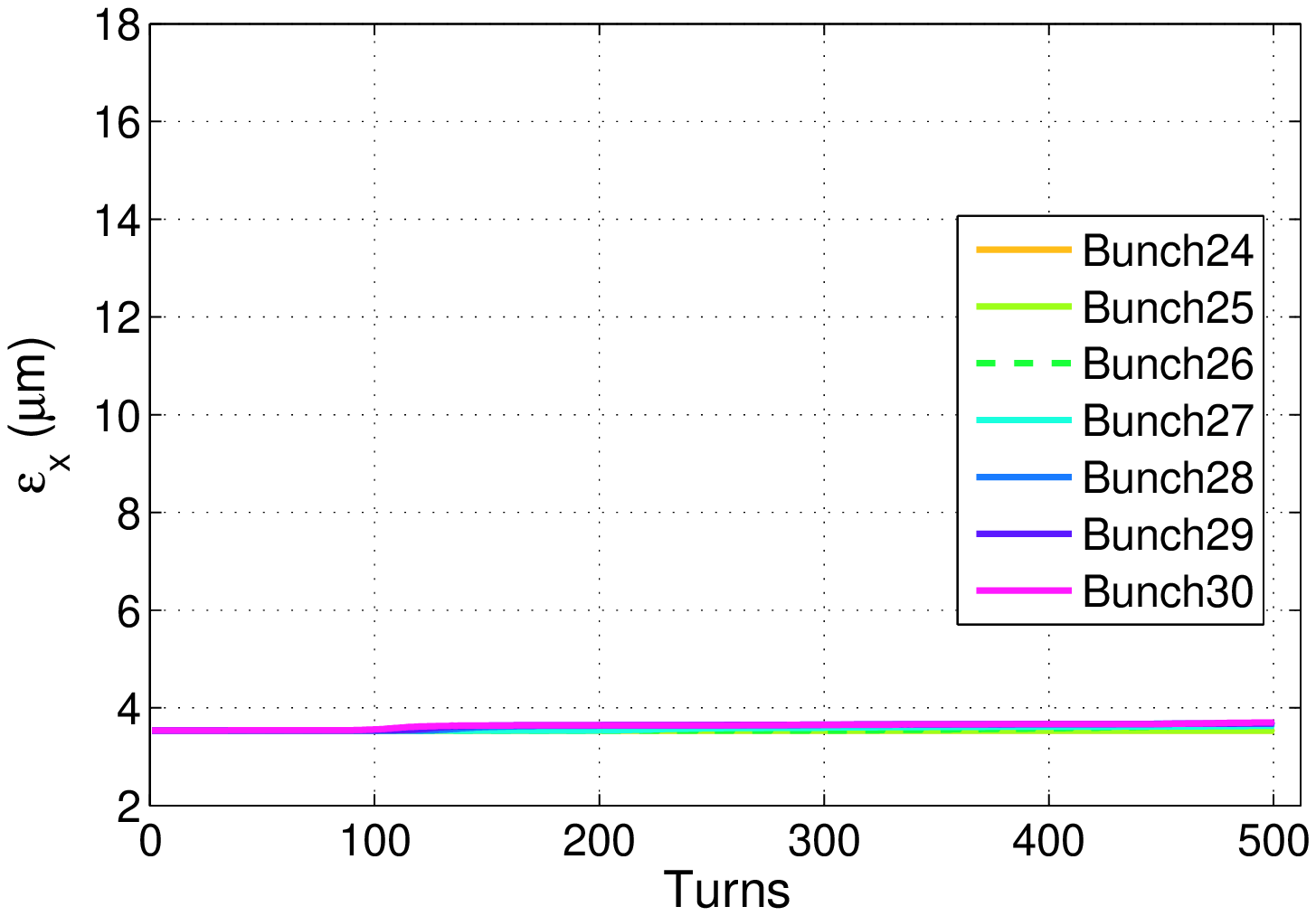}
    \includegraphics[trim= 0 0 0 0, clip,width=0.42\textwidth]{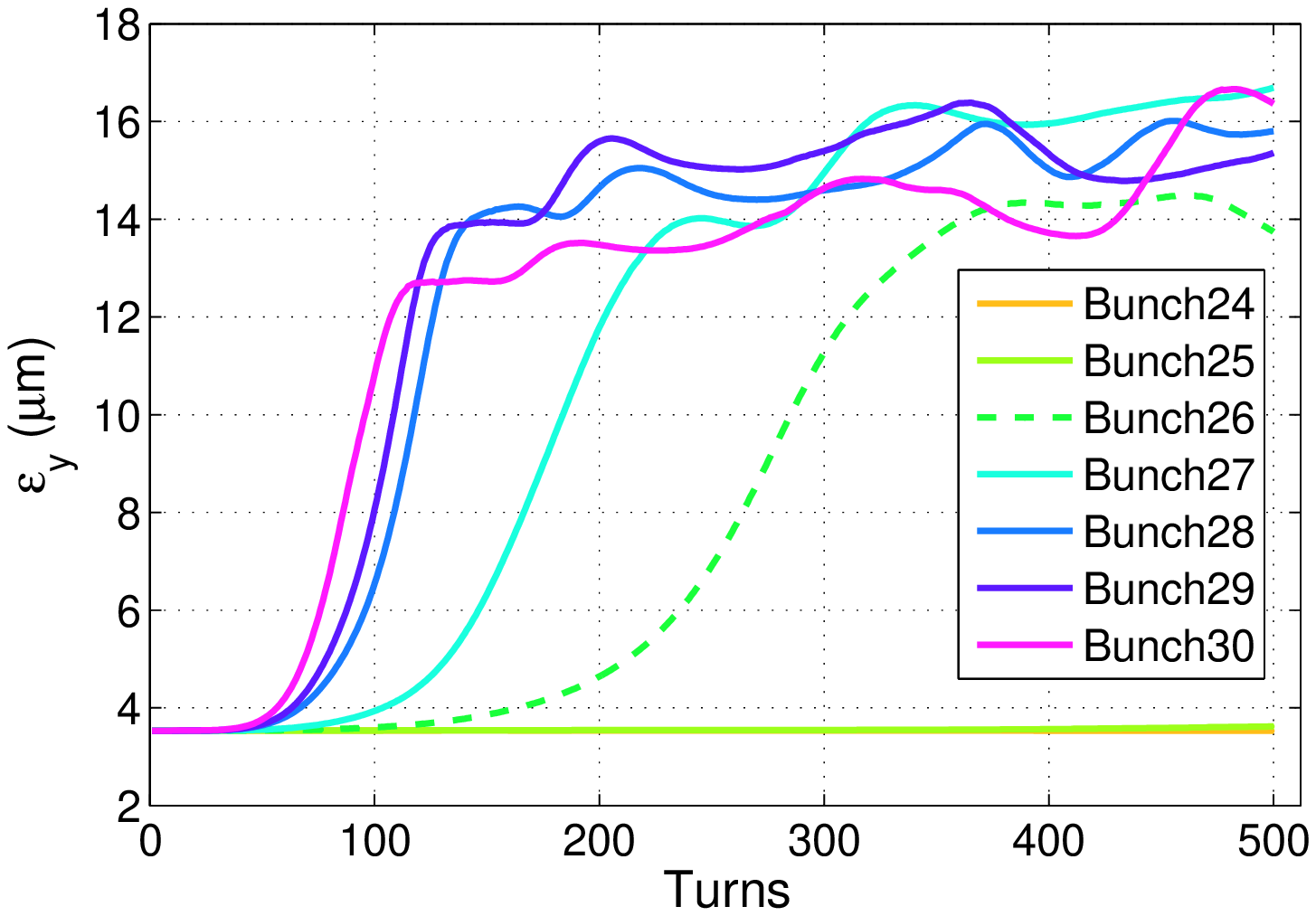}
    \caption{Evolution of the horizontal (top) and vertical (bottom) emittance along 500 turns for a few selected bunches in the center of the bunch train.} 
    \label{FIG:HEADTAILsimulation}
\end{figure}

\begin{figure}[h]
    \centering
    \includegraphics[trim= 0 0 0 0, clip,width=0.48\textwidth]{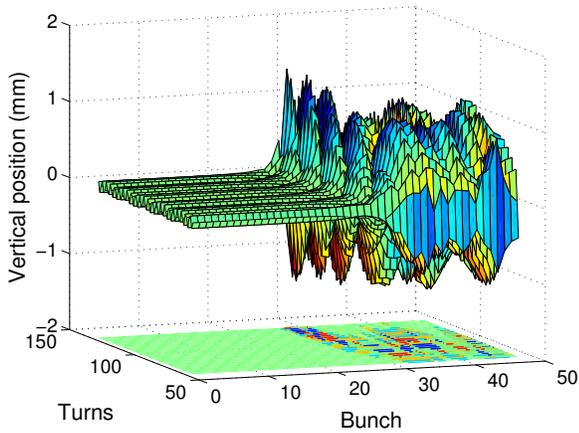}
    \caption{Bunch-by-bunch oscillations in the vertical plane as obtained with HEADTAIL. For better comparison, only 70 turns during the instability (from turn 50 to turn 120) are shown.} 
    \label{FIG:BbBOscillationsHT}
\end{figure}

The above described experimental observations were studied by combining PyECLOUD~\cite{GiadarolECLOUD12} and HEADTAIL~\cite{HEADTAIL} simulations. As first step the electron cloud build-up in the LHC dipoles is simulated with PyECLOUD for an r.m.s.~bunch length of 12\,cm  and equal bunch intensities of  $N=1.0\times10^{11}$\,p/b along the train\footnote{It was not possible to retrieve the actual bunch-by-bunch intensity variation from the data logging during the injection tests due to the short time of circulating beam before the dump.}. Figure\,\ref{FIG:PyECLOUDbuildup} shows the obtained electron density per unit length along the bunch train for a secondary electron yield of $\delta_\text{max}\!=\!2.1$. The simulation was initialized with a uniform distribution using the same number of primary electrons as in the simulations for determining the evolution of the SEY discussed above \cite{GrumoloEvian2011}. The importance of the number of primary electrons and their spatial distribution for the onset of the e-cloud instability will be discussed in more detail in the next section. The red markers indicate the time steps in the simulation just before the bunch passages. The electron distributions sampled at these points are used as input for a set of 48 HEADTAIL simulations, one for each bunch. 
\begin{figure}[h]
    \centering
    \includegraphics[trim= 0 0 0 0, clip,width=0.45\textwidth]{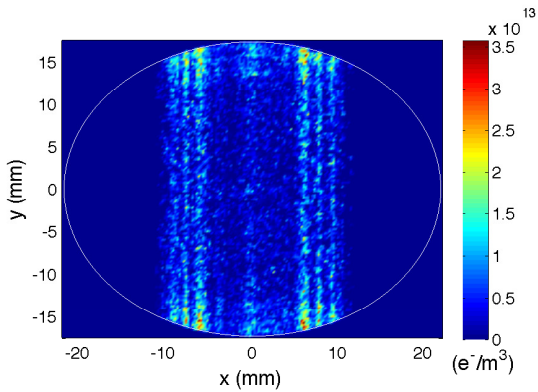}
    \caption{Electron distribution before the passage of the last bunch of the train as obtained from the PyECLOUD simulation. Note that the cross section of the LHC beam screen is approximated as ellipse.} 
    \label{FIG:ElectronDistributionBunch48}
\end{figure}
Figure~\ref{FIG:ElectronDistributionBunch48} shows for example the electron distribution before the entrance of the last bunch of the train. Note the accumulation of electrons in two stripes, as typically obtained for the build-up with 25 ns bunch spacing in a geometry as the LHC beam screen when the electrons move in a strong dipolar magnetic field. In HEADTAIL the electron cloud is represented by thin slices lumped at several accelerator sections and the electron motion is frozen in the horizontal plane in order to  account for the effect of the dipole magnetic field. At each electron cloud section the electron proton interaction is computed consecutively for longitudinal bunch slices. After a complete bunch passage the electron cloud is reset to the initial distribution for the next interaction section. Figure~\ref{FIG:HEADTAILsimulation} shows the evolution of the transverse emittance for a few selected bunches in the middle of the bunch train obtained with HEADTAIL for the simulation of 500 turns at the LHC injection plateau using the initial electron distribution as described above. 
No instability or coherent excitation of the horizontal bunch motion is observed, as was expected since the electron cloud is located in dipole regions. On the other hand the electron cloud drives a single bunch instability in the vertical plane. This instability can be observed in the form of an exponential emittance growth, as in the case studied here for all bunches after bunch number 25. Figure\,\ref{FIG:BbBOscillationsHT} shows the bunch-by-bunch oscillations in the vertical plane during the instability (from turn 50 to turn 120) as obtained from HEADTAIL. 
 The simulation is in good agreement with the experimental observations during the first injection of 48 bunches in August with the transverse damper on, where the second half of the bunch train is unstable mainly in the vertical plane (cf.~Fig.\,\ref{FIG:ADToscillations}), as the damper is suppressing horizontal coupled bunch instabilities. Horizontal instabilities are not observed in the simulation, since the coupling between bunches is not taken into account but each bunch is treated by an independent HEADTAIL simulation. To include this coupling would require a combined simulation of PyECLOUD and HEADTAIL, which is envisaged to be implemented at a later stage. This will allow to reproduce better the instability observed during the injection without transverse damper.
 
\section{Sensitivity studies} 

It was shown in the previous section that the instability observed at LHC injection can be reproduced with the presented simulation procedure. In the following the sensitivity of the simulation result to the number and distribution of the primary electrons will be studied. Furthermore, it is observed that the electron cloud is formed mainly in two symmetric vertical ``stripes'' around the beam center (cf.\,Fig.\,\ref{FIG:ElectronDistributionBunch48}) and only a small number of electrons is concentrated in the central part of the chamber. Thus it is interesting to assess which part of the electron cloud distribution is mainly responsible for driving the beam unstable. 

\subsection{Electron density}

\begin{figure}[h!]
    \centering
    \includegraphics[trim= 0 -9 0 0, clip,width=0.43\textwidth]{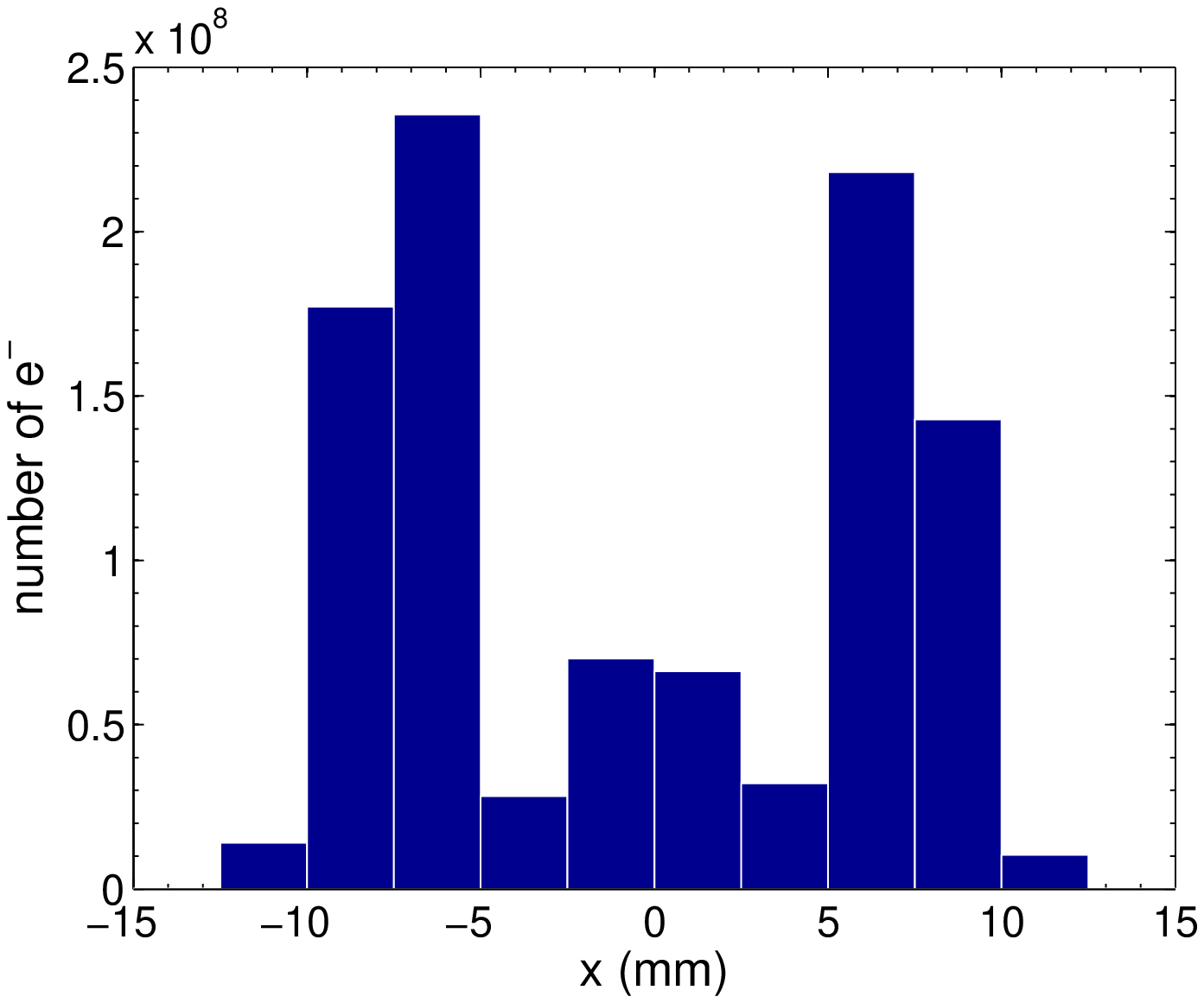}
    \includegraphics[trim= 0 -9 0 0, clip,width=0.43\textwidth]{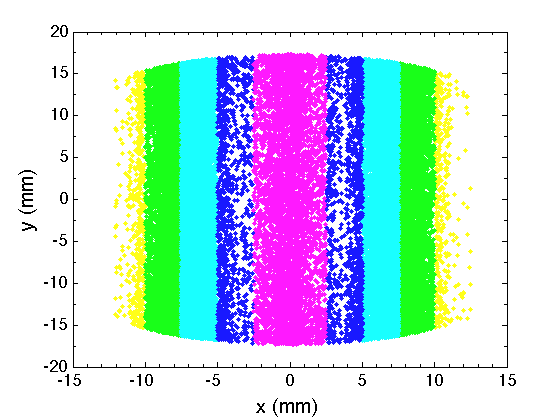}    
    \includegraphics[trim= 0 -9 0 0, clip,width=0.43\textwidth]{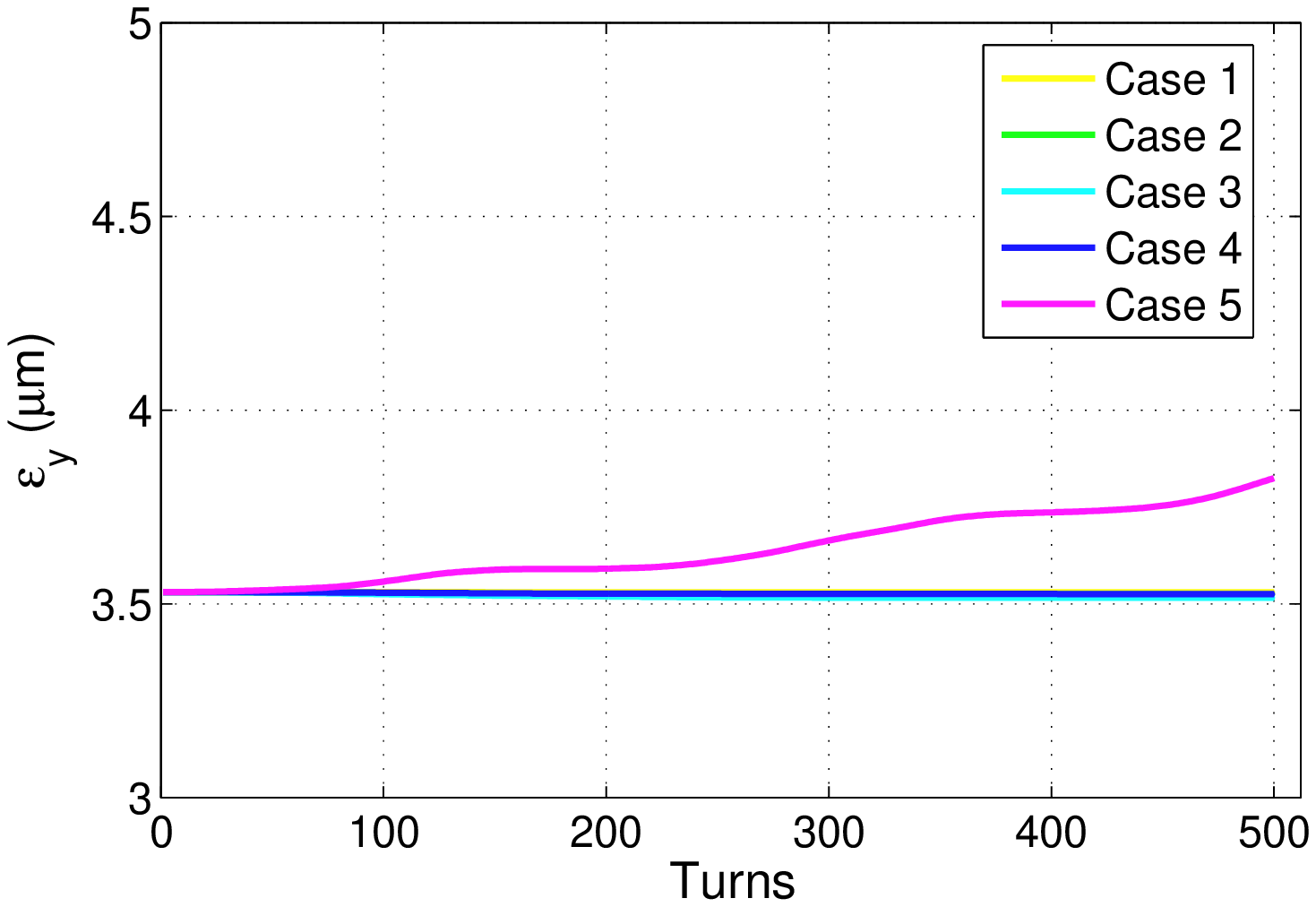}
    \includegraphics[trim= 0 -9 0 0, clip,width=0.43\textwidth]{VerticalEmittance.eps}
    \caption{From top to bottom: Histogram of the horizontal electron distribution; transverse electron distribution divided into colored regions; evolution of the vertical emittance for the interaction with the electrons of the respective colored area;  evolution of the vertical emittance for different horizontal cuts of the electron distribution.}
    \label{FIG:CentralDensity}
\vspace{-1cm}    
\end{figure}

Figure\,\ref{FIG:CentralDensity} shows an example for a bunch in the middle of the train, which encounters an electron density which is just above the instability threshold. The top graph shows a histogram of the horizontal electron distribution before the bunch arrival as obtained from the PyECLOUD simulation and used as input for the HEADTAIL simulation. Note again 
that most of the electrons are concentrated in the two stripes far away from the beam (the $3\,\sigma$ beam envelope covers roughly the two central bins). The graph below shows for the same case the actual positions of the macro-electrons in the transverse plane. The macro-particle distribution is divided here into five regions as indicated by the color code (corresponding to the bins in the histogram). In the following, the contribution from these five regions to the vertical instability is studied. First, each of these regions is considered independently and used as input for individual HEADTAIL simulations. The evolution of the vertical emittance for these five cases is shown for 500 turns in the third graph of Fig.\,\ref{FIG:CentralDensity}. In the case studied here, only the central part of the electron distribution is able to drive the beam unstable. The bottom graph shows the vertical emittance evolution in case all electrons enclosed by the respective colored areas have been removed, i.e.~cutting the electron distribution at the inner borders of the colored regions. The instability appears only when the electrons in the central part of the distribution are taken into account. It follows that for the typical electron cloud distributions encountered in the simulations of the LHC dipole regions, it is mainly the central electron density which determines the onset of the instability. Therefore the instability threshold can be inferred roughly from the central electron density. Only in cases where the central electron density is \emph{very} small compared to the density in the stripes this approximation may not hold. 

In the following, the central density of the electron distribution is thus used to determine the instability threshold. It should be noted that the instability thresholds found in this manner are consistent with thresholds previously found in HEADTAIL simulations assuming a uniform electron distribution before the bunch passage, which is around $1\times10^{12}\,e^-/$m$^3$ \cite{KLi2011}. 


\subsection{Dependence on number of primary electrons}

\begin{figure}[h]
    \centering
    \includegraphics[trim= 0 -4 0 -4, clip,width=0.43\textwidth]{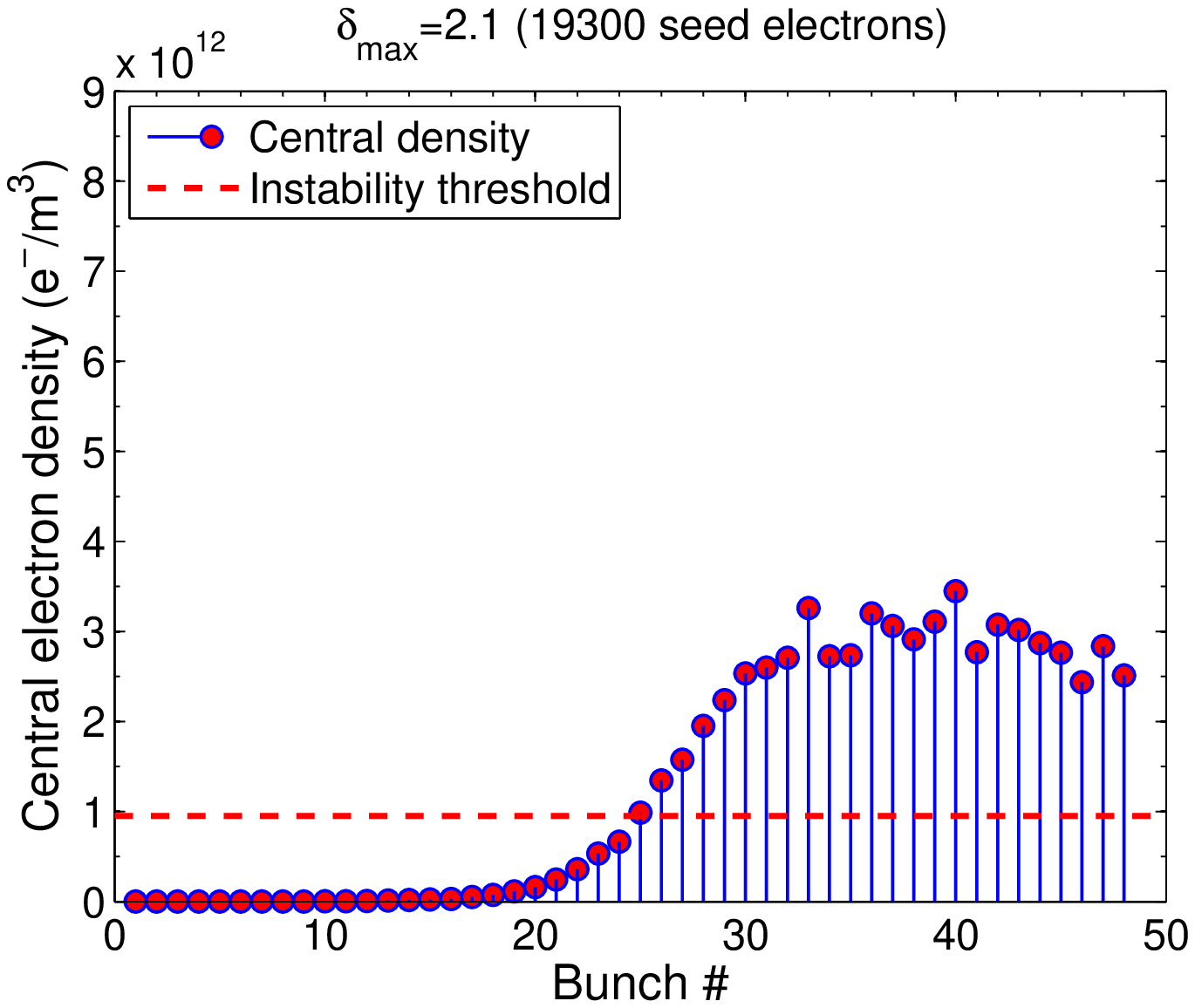}
    \includegraphics[trim= 0 -4 0 -4, clip,width=0.43\textwidth]{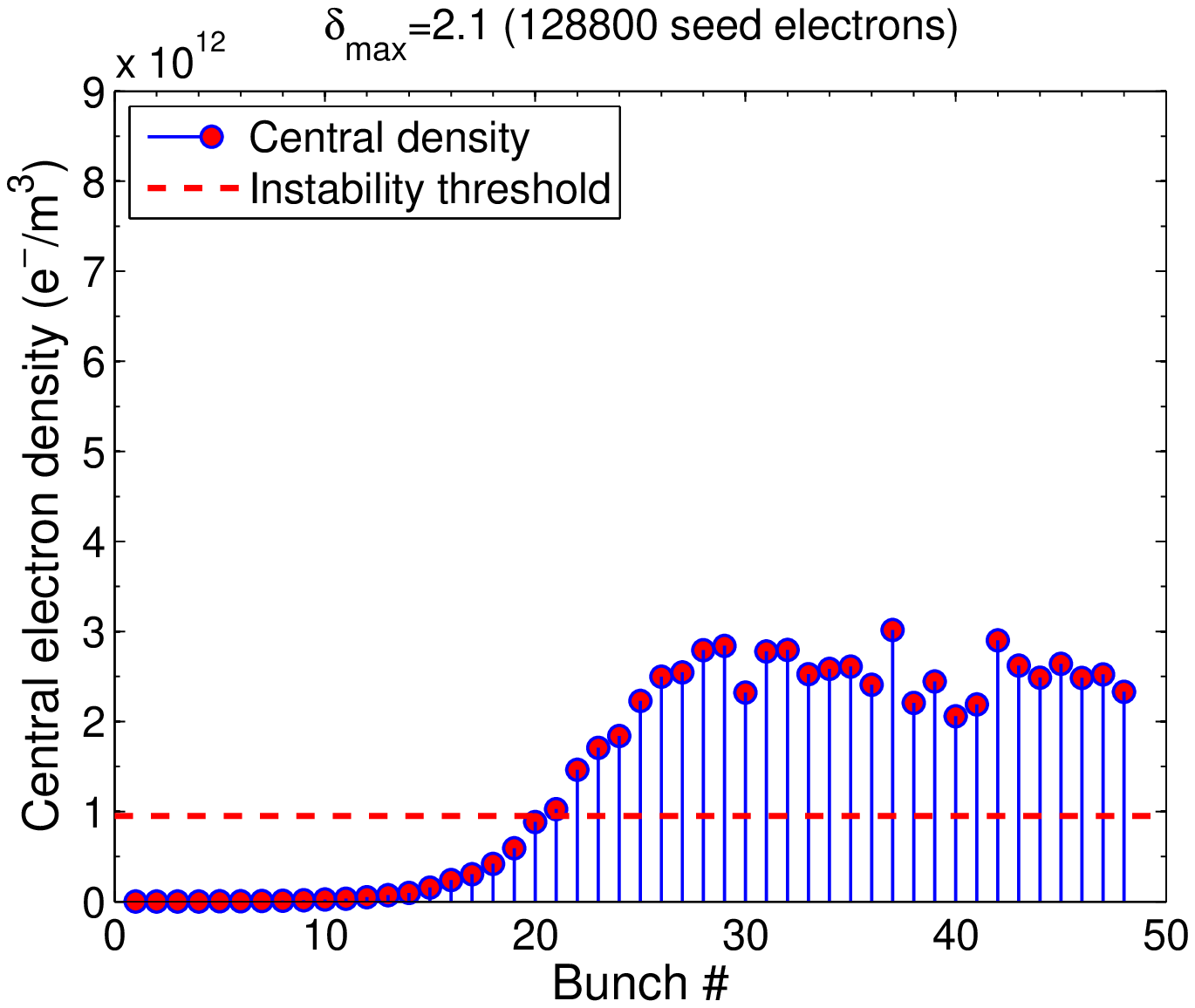}    
    \caption{Comparison of the central electron density along the bunch train for a number of seeding electrons comparable to the static pressure in the LHC dipole sections (top) and for a larger number of seeding electrons (bottom).} 
    \label{FIG:DependenceOnNumberOfSeeds}
\end{figure}

One of the main uncertainties of the electron cloud build-up simulations for the LHC injection energy is the number of primary (seed) electrons. Since the synchrotron radiation at 450\,GeV is inefficient to generate photo electrons, it is assumed that the primary electrons are created by rest gas ionization. Therefore the number of primary electrons depends on the pressure. However the static pressure without beam is much lower compared to the pressure levels measured  after beam injection. Figure\,\ref{FIG:DependenceOnNumberOfSeeds} shows the central density along the bunch train of 48 bunches for $\delta_\text{max}=2.1$ and two different cases: for a small number of seed electrons equivalent to the static pressure in the LHC cold sections (top) and for a large number of seed electrons (bottom). The dashed red line shows the central electron density at the instability threshold. As expected, it takes more bunch passages to build up the electron cloud beyond the instability threshold for a smaller number of primary electrons. The onset of the instability along the bunch train depends thus on the pressure level assumed in the build-up simulation. For completeness it should be emphasized that the saturation level of the central density is very similar in both cases (as it depends mostly on $\delta_\text{max}$). 

\subsection{Dependence on primary electron distribution}

\begin{figure*}[t]
    \centering
    \includegraphics[trim= 1 0 0 -8, clip,width=0.33\textwidth]{SEY2p1_30nT_1p0unif.eps}
    \includegraphics[trim= 1 0 0 -8, clip,width=0.33\textwidth]{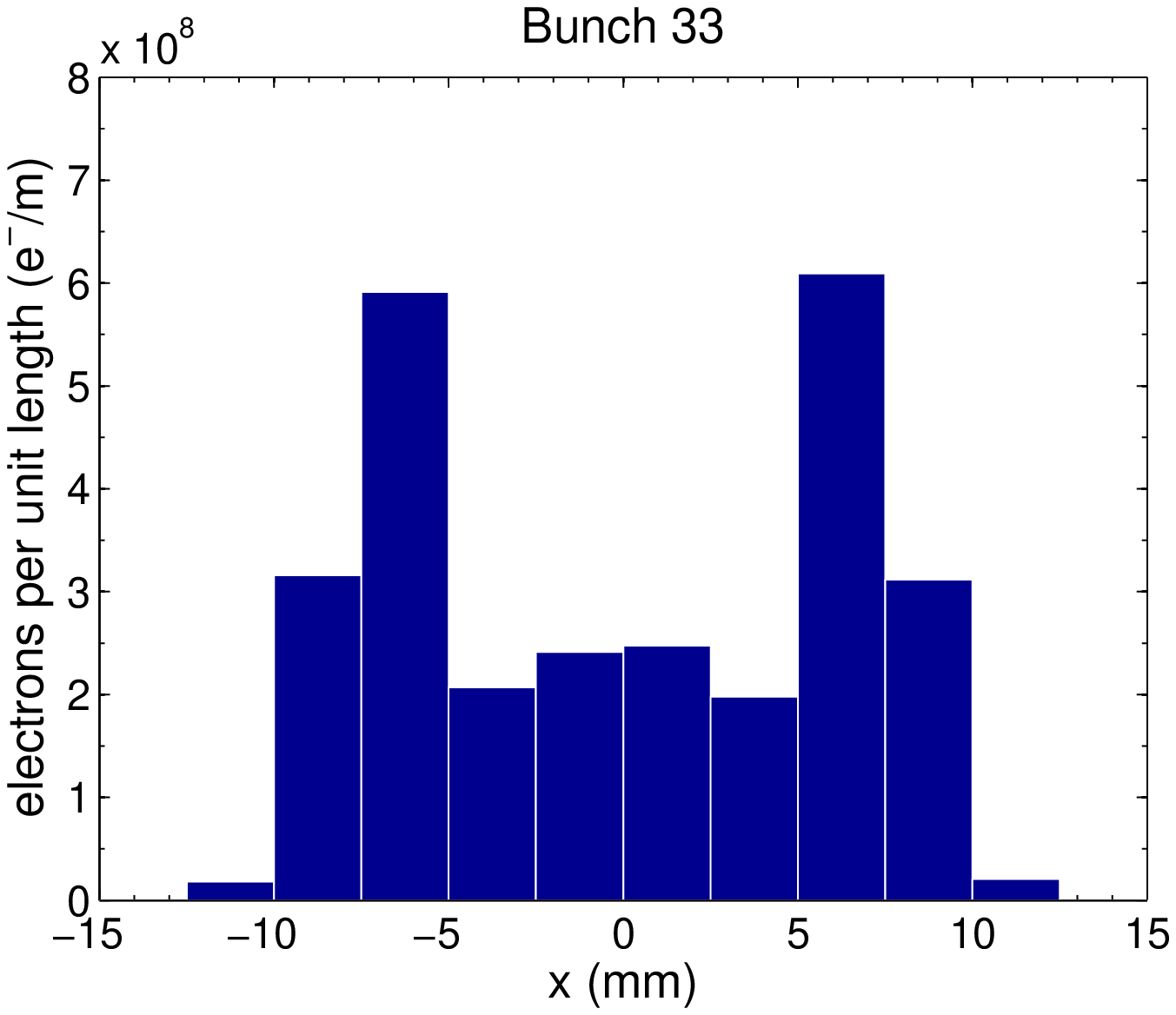}    
    \includegraphics[trim= 1 0 0 -8, clip,width=0.33\textwidth]{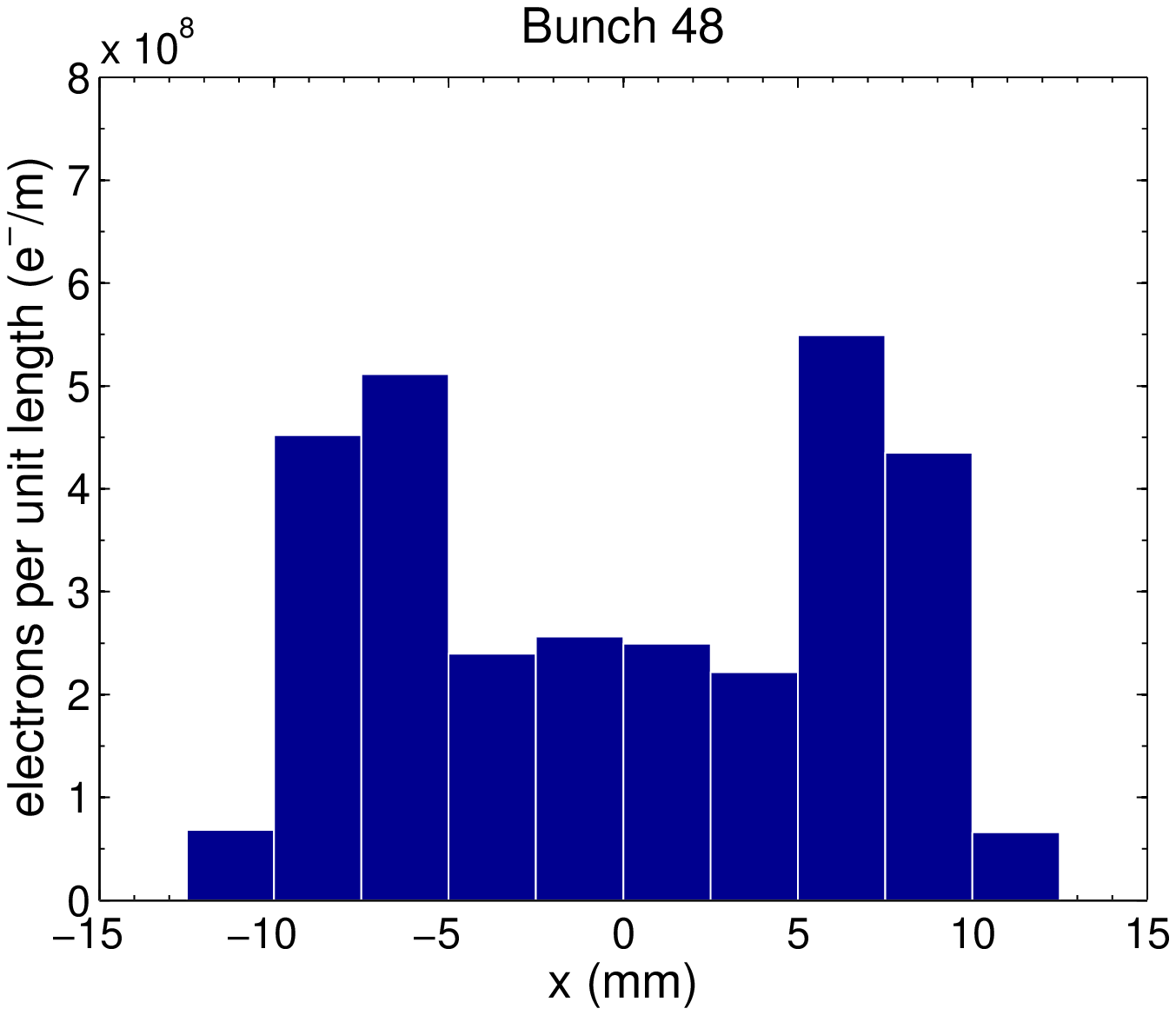}        
    \caption{Example for an electron cloud build-up simulation assuming a uniform transverse distribution of the primary electrons all across the vacuum chamber: central density along the bunch train (left), horizontal distribution for bunch 33 (middle) and for the last bunch (right).} 
    \label{FIG:Uniform}
\end{figure*}

\begin{figure*}[t]
    \centering
    \includegraphics[trim= 1 0 0 -8, clip,width=0.33\textwidth]{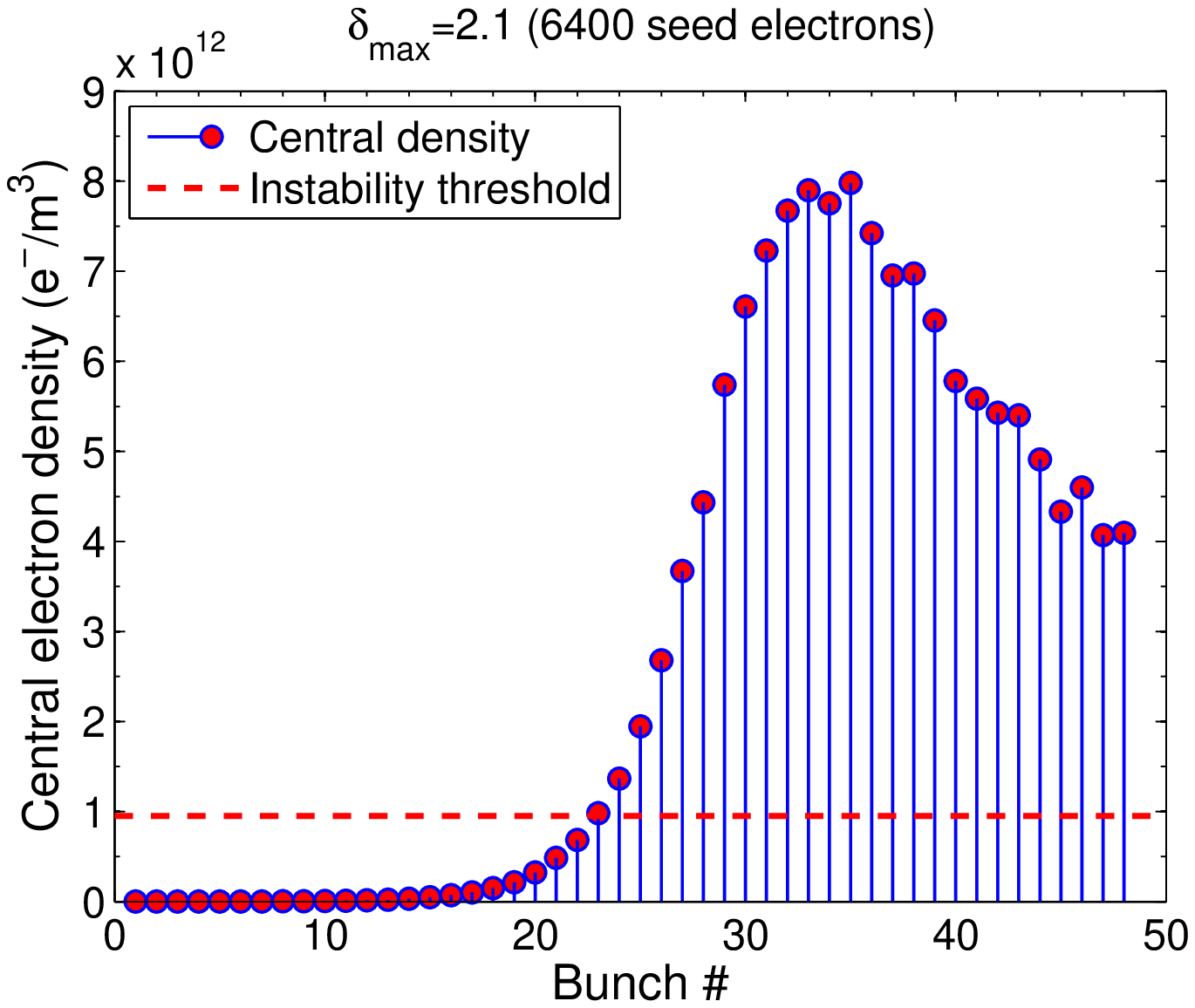}
    \includegraphics[trim= 1 0 0 -8, clip,width=0.33\textwidth]{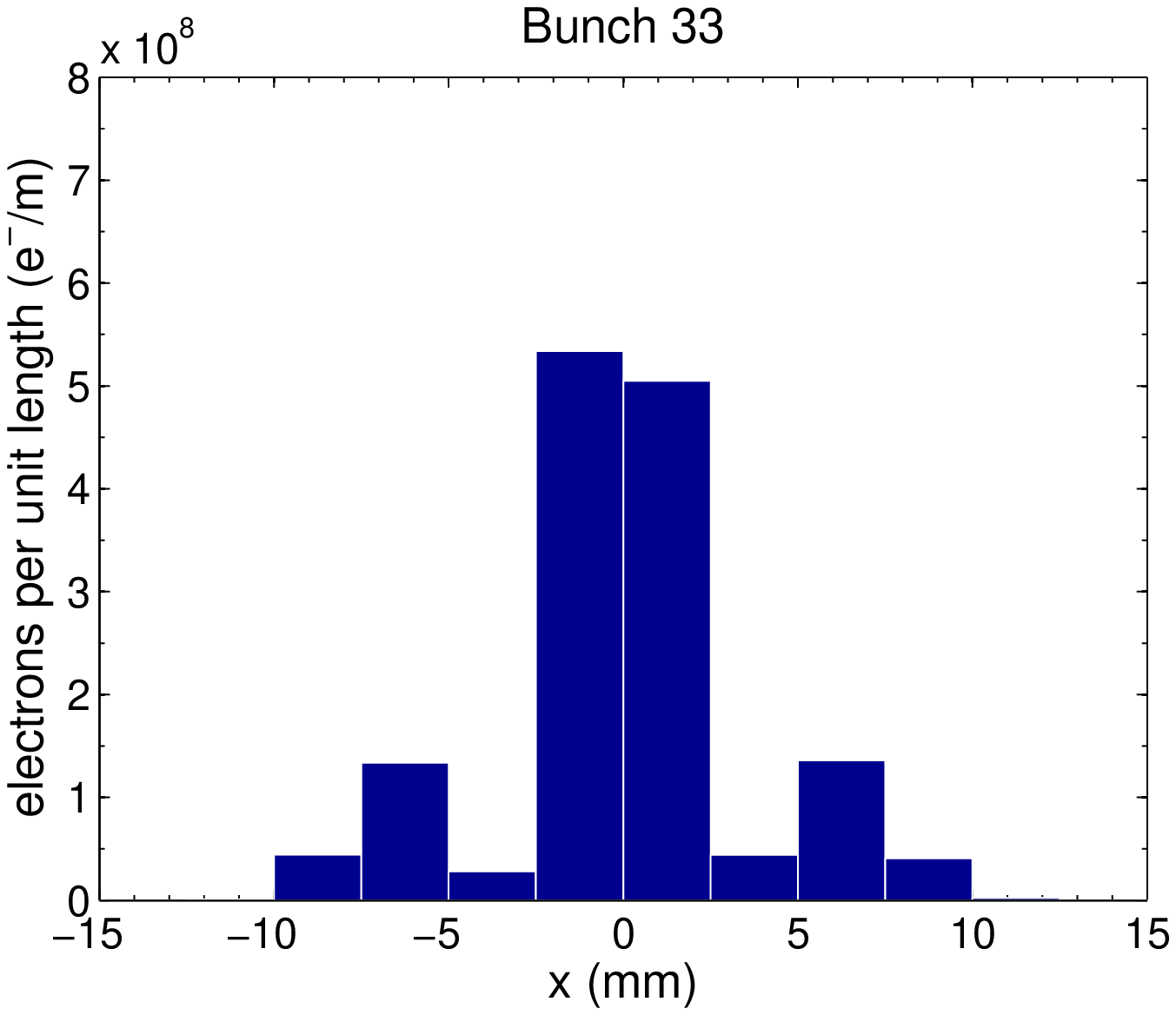}    
    \includegraphics[trim= 1 0 0 -8, clip,width=0.33\textwidth]{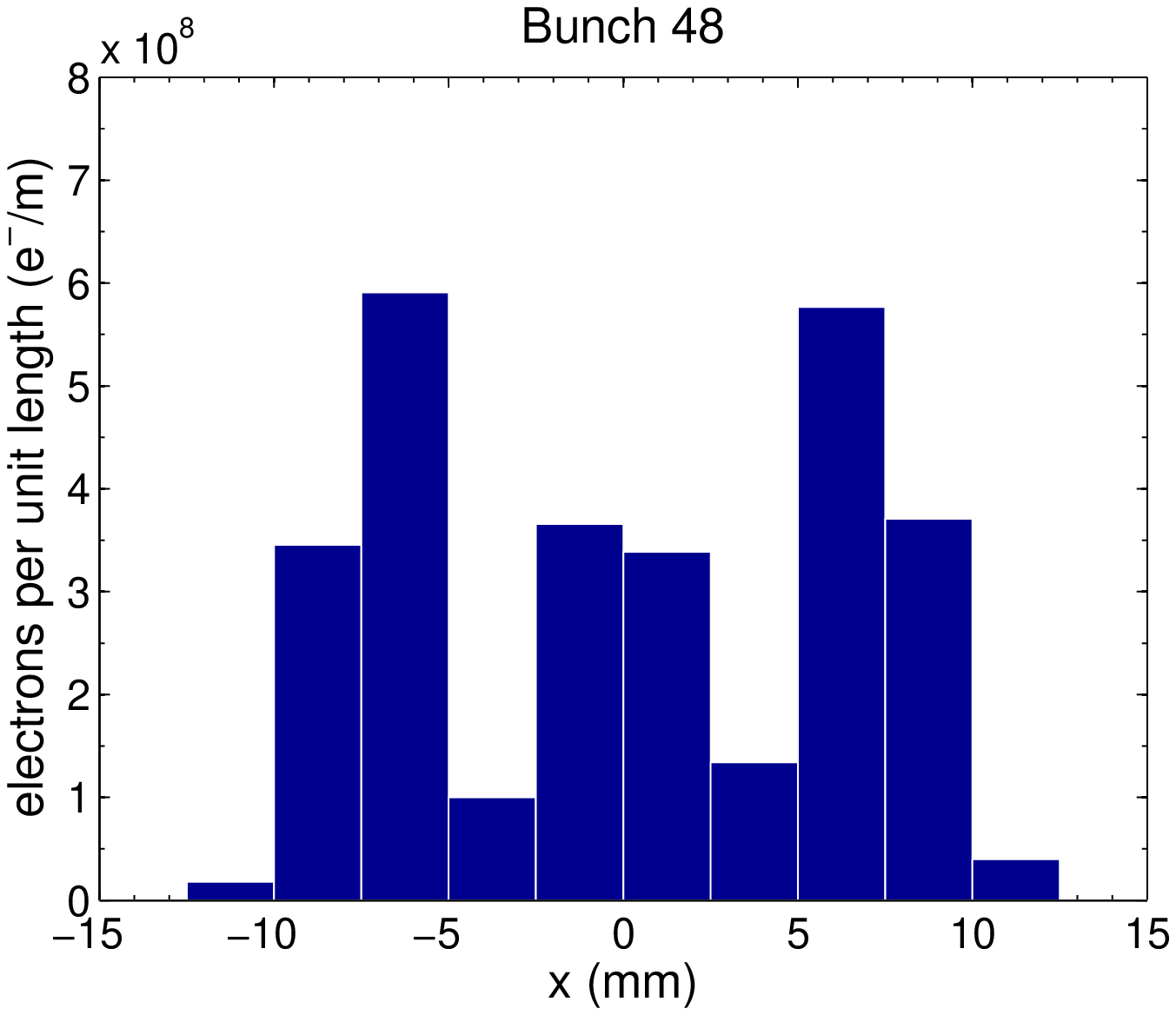}        
    \caption{Example for an electron cloud build-up simulation assuming a Gaussian distribution of the primary electrons on top of a 10\% uniform background all across the vacuum chamber: central density along the bunch train (left), horizontal distribution for bunch 33 (middle) and for the last bunch (right).} 
    \label{FIG:GaussianPlusBackground}
\end{figure*}

In addition to the number of primary electrons, also their distribution is not well known in the case of the LHC at injection energy. In order to reproduce the experimentally observed vertical stripes of the electron distribution in saturation, primary electrons have to be assumed across the entire cross section of the vacuum chamber (since the electrons are bound by the vertical magnetic field lines). Therefore a uniform distribution of the primary electrons can be optionally  used in PyECLOUD simulations. Figure\,\ref{FIG:Uniform} shows an example for a uniform distribution of the seed electrons together with the histogram of the horizontal electron distribution for two selected bunches (a bunch in the second half of the train where the saturation level is reached, and the last bunch). The central density is very similar in the two cases, but the stripes build-up further outside towards the end of the bunch train. 

Considering that the generation of the primary electrons is caused by rest gas ionization, it can be argued that the primary electrons should follow a Gaussian distribution similar to the proton beam. In this case the formation of the stripes can be achieved by adding about 10\% of the total number of electrons in the form of a uniform background all across the vacuum chamber. Figure\,\ref{FIG:GaussianPlusBackground} shows the evolution of the central electron density along the bunch train assuming such a distribution of seed electrons together with the horizontal electron distribution for selected bunches as in Fig.\,\ref{FIG:Uniform}. Here a smaller number of seed electrons was chosen in order to obtain the instability onset roughly at the same bunch along the bunch train as for the case of the uniform distribution of seed electrons. However, the evolution of the electron distribution along the bunch train is quite different for the two cases. In comparison to the case of the uniform distribution, the central density reaches much higher peak values  in the saturated part of the train and it takes longer to develop the stripes in the outer part of the vacuum chamber assuming the Gaussian distribution. This can be understood intuitively, since the center of the chamber is seeded with a larger number of electrons while the outer regions of the vacuum chamber are seeded with a comparably smaller number of electrons. 

\section{Compatible parameter space}

\begin{figure}[h]
    \centering
    \includegraphics[trim= 0 0 0 0, clip,width=0.45\textwidth]{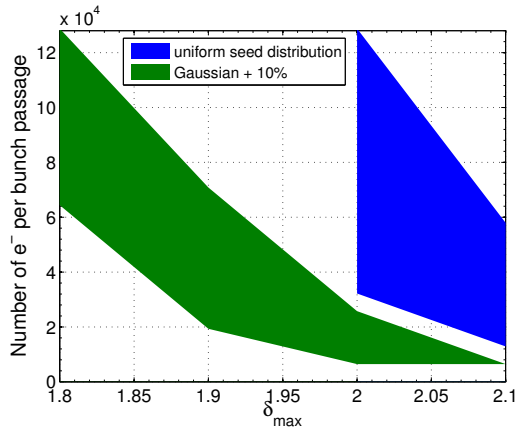} 
    \caption{Range of parameters reproducing the observed instability in the LHC, as indicated by the colored areas assuming a uniform primary electron distribution (blue) and the Gaussian distribution on top of a 10\% background (green).} 
    \label{FIG:CompatibleSolutions}
\end{figure}

In fact the number and distribution of the primary electrons are among the biggest uncertainties in the present understanding of the electron cloud build-up in the LHC dipoles at injection energy. On the other hand, the estimation of $\delta_\text{max}$ from the reproduction of the measured heat load in the LHC beam screens with PyECLOUD \cite{GrumoloChamonix2012} is not so sensitive to the assumptions on the seed electrons if the bunch train is long and the electron saturation level is reached for many bunches. The estimation of $\delta_\text{max}$ for the case studied here was not done for the same time as the observations of the injections of the 48 bunches and therefore might be slightly smaller than $\delta_\text{max}=2.1$. In the following, the available parameter space is thus scanned for compatible solutions reproducing the observed instability at LHC injection in August 2011. Figure\,\ref{FIG:CompatibleSolutions} shows compatible solutions, i.e.~cases where the onset of the instability is close to bunch 25 (between bunch 23 and bunch 27), for a range of possible values for $\delta_\text{max}$ and for the number of primary electrons per bunch passage. The two colored regions correspond to the cases of a purely uniform distribution of the primary electrons and a Gaussian distribution with a 10\% uniform background, respectively. As expected, higher pressure levels (or equivalently a larger number of primary electrons) and a larger secondary electron yield is needed to reproduce the observations in the case of a uniform electron distribution, due to the smaller central electron density compared to the case with a Gaussian electron distribution. Note that the range of the number of seed electrons per bunch passage explored in this study corresponds to room-temperature equivalent pressures between 10\,nTorr and 200\,nTorr (the static pressure in the LHC cold sections is around 32\,nTorr room-temperature equivalent). 
Further studies will be needed in the future in order to better understand the mechanism of the primary electron production and their distribution at LHC injection energy.

\section{CONCLUSIONS AND OUTLOOK}

Fast instabilities were observed during the first two attempts of injecting bunch trains of 48 bunches with the nominal 25\,ns bunch spacing in August 2011. 
The onset of vertical single bunch instability can be reproduced in good agreement by electron cloud simulations using a combination of PyECLOUD for the build-up and HEADTAIL for the beam dynamics part. The simulations are based on values of $\delta_\text{max}$ as estimated from the measured heat load data in the LHC using PyECLOUD. The presented studies based on a combination of PyECLOUD with the HEADTAIL code can therefore be considered as consistency check of the current model of the electron-cloud effects on the LHC flat bottom. The simulations show a vertical instability, fast emittance growth and strong losses, similar to observations in the LHC. The instability onset depends strongly on the central electron density seen by the beam, which itself depends on the initial conditions assumed for the build-up, namely the number of primary electrons and their distribution. Future studies should aim at improving the understanding of the mechanism responsible for the generation of the primary electrons. This will become important also for the estimation of $\delta_\text{max}$ from the measured heat load close to the threshold of the electron cloud build-up, despite the fact that the beam is not expected to suffer from the instability any more. Apart from this, it would be interesting to include the coupling between the bunches due to the electron cloud in the simulations. This would require a self consistent model of the electron cloud effects like a combination of PyECLOUD and HEADTAIL in one big simulation for all bunches, which would then allow to study the observed horizontal coupled bunch instability. Finally it might be interesting to include the effect of the transverse feedback in HEADTAIL, as this would help to estimate better the required settings of chromaticity and octupoles to stabilize the beam in a strong electron-cloud regime.

\section{ACKNOWLEDGEMENTS}
The authors would like to thank D.\,Valuch for help extracting and analyzing the turn-by-turn data from the damper pick-ups and feedback setup, G.\,Arduini for valuable discussion, and K.\,Li for support for the HEADTAIL simulations.


\end{document}